\documentclass[%
 reprint,
 amsmath,amssymb,
 aps,
]{revtex4-2}

\usepackage{graphicx}
\usepackage{dcolumn}
\usepackage{bm}

\usepackage[utf8]{inputenc}
\usepackage{amsmath}
\usepackage{graphicx}
\usepackage{booktabs,siunitx}
\usepackage{hyperref}
\usepackage{amsfonts}
\usepackage{bbm}
\usepackage{bigints}
\usepackage[utf8]{inputenc}
\usepackage{amssymb}
\usepackage{pifont}

\usepackage{xtab,afterpage}
\usepackage{setspace}

\usepackage{color,soul}
\usepackage{xcolor}

\begin{document}

\title{\textit{Ab Initio} Modeling of Thermal Transport through van der Waals Materials}

\author{Sara Fiore}
 \altaffiliation{Corresponding author: safiore@iis.ee.ethz.ch}
\author{Mathieu Luisier}%
\affiliation{Integrated Systems Laboratory, ETH Zurich, 8092 Zurich, Switzerland
}%

\begin{abstract}
An advanced modeling approach is presented to shed light on the thermal transport properties of van der Waals materials (vdWMs) composed of single-layer transition metal dichalcogenides (TMDs) stacked on top of each other with a total or partial overlap only in the middle region. It relies on the calculation of dynamical matrices from first-principle and on their usage in a phonon quantum transport simulator.  
We observe that vibrations are transferred microscopically from one layer to the other along the overlap region which acts as a filter selecting out the states that can pass through it. Our work emphasizes the possibility of engineering  heat flows at the nanoscale by carefully selecting the TMD monolayers that compose vdWMs.
\end{abstract}

\maketitle

\section{Introduction}
During the last 50 years electronic devices have undergone a constant miniaturization of their dimensions following  Moore's scaling law \cite{THOMPSON200620}.
While the integration of always smaller components has paved the way for  enhanced  functionabilities, several obstacles have recently emerged that could compromise the benefit of further size reductions \cite{ball2012computer}. In particular, modern transistors already operate at length scale of the same order as the electron and phonon mean free paths \cite{inproceedings} where discrete scattering events and geometrical variations have a profound impact on their behaviour. The increased power density of these devices combined with high thermal dissipation and dramatic peak temperatures, severely limit their reliability, performance and lifetime \cite{4675673},\cite{1705144}. Another critical issue is self- or Joule-heating, a phenomenon caused by the phonon emission of  high energy electrons \cite{Sood_2018}. 
Instead of eliminating thermal fluctuations several research efforts have tried to take advantage of nanoscale systems to recycle the waste heat through Seebeck's effect \cite{Biele_2017},\cite{he2018thermoelectric}. 
Designing such thermoelectric devices or next-generation logic switch will only be possible if both their electronic and thermal properties are kept under control and their interactions well understood. 
Physics-based materials and device modeling can be of great help for that purpose, especially if combined with experiments in an holistic way.
So far, the focus has been mainly set on the electronic characteristics of nanostructures \cite{mimura1980new},\cite{faist1994quantum},\cite{yablonovitch1993photonic},\cite{CHENG2014348},\cite{ionescu2011tunnel},\cite{dewey2011fabrication},\cite{ghosh2013monolayer},\cite{lam2013device},\cite{britnell2012field} and much less on their thermal behaviour \cite{Maldovan2013SoundAH},\cite{34d524ac883843cc8d18933b99521d10},\cite{rodgers2004prediction},\cite{Wan}.

\begin{figure}[h!]
    \centering
    \includegraphics[width=\linewidth]{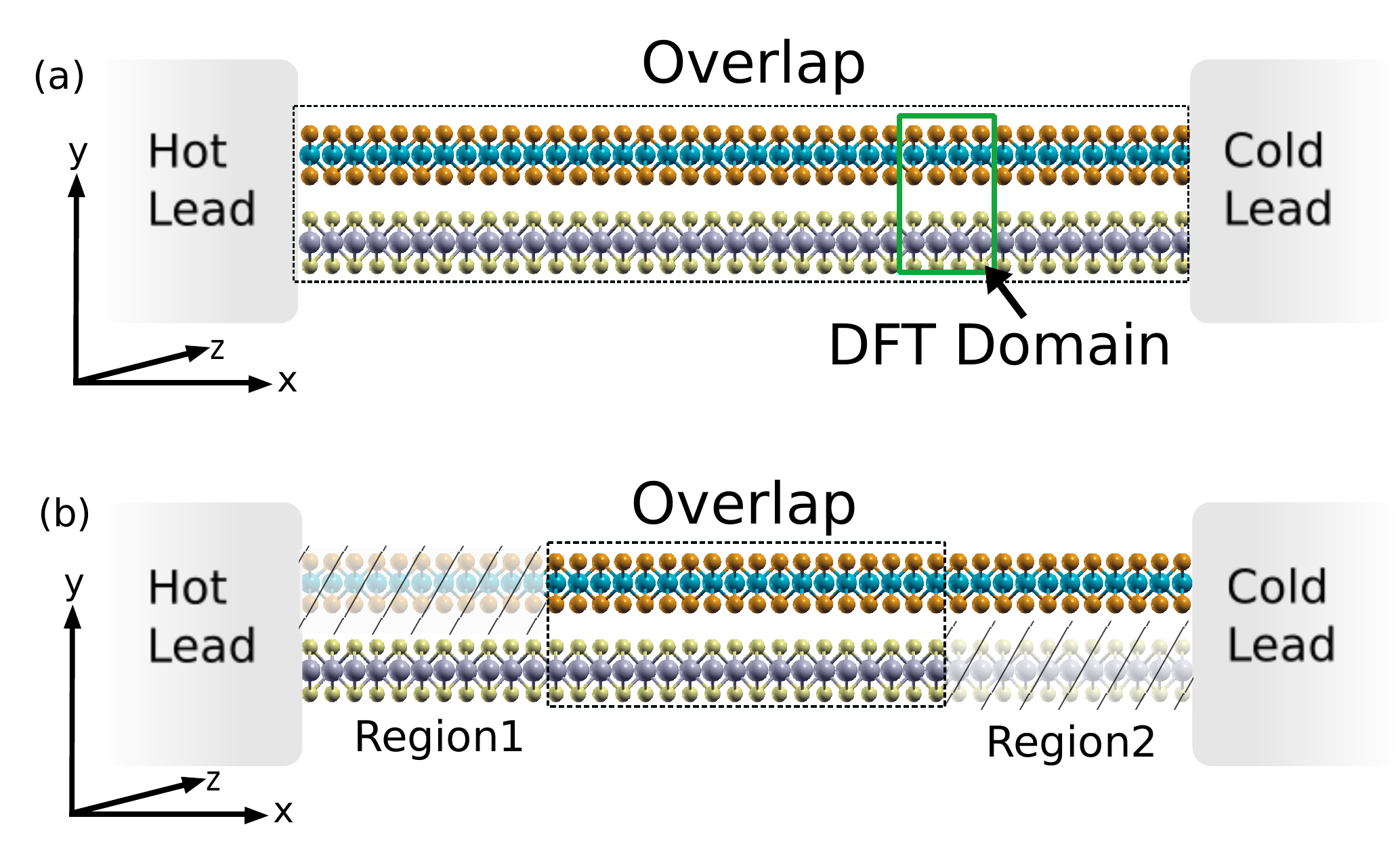}
    
    \caption{(a) vdWM made of two TMDs with complete overlap. Such structures are referred to as TO channels. (b) Same as (a), but with a partial overlap of both TMDs in the middle and two regions made of single TMDs. Such structures are referred to as PO channels. \label{fig:vdWHs}}
\end{figure}

To fill this gap and gain insight into nanoscale thermal management, we propose an original approach to model thermal transport and apply it to artificial materials called van der Waals Materials (vdWMs).
Due to their ultimate thickness, van der Waals inter-layer coupling, intra-layer covalent bonds and surface free of dangling bonds, these compounds are expected to exhibit unique features \cite{akinwande2014two},\cite{osada2009exfoliated},\cite{miro2014atlas},\cite{Kumar_2016}. 
Since the breakthrough exfoliation of graphene in 2004 \cite{Novoselov_2004},\cite{Novoselov_2005} more than 1800 layered materials have been predicted to exist and some of them have already been intensively investigated \cite{berger2019nanoengineering}, \cite{materialcloud}. In the 3D parents of 2D materials, strong covalent bonds hold together the crystalline structure in the in-plane direction, while van der Waals forces act along the out-of-plane axis connecting different layers.
Hence, monolayers and few layers structures can be isolated and/or stacked on top of each other to form carefully designed van der Waals materials. Each 2-D material can be considered as a Lego brick that can be assembled with other building blocks.
Several studies have been carried out on vdWMs, highlighting their potential as active region of future electronic (vertical field-effect transistors \cite{FET_vdWH} and P-N diodes \cite{PN_vdWH}), spintronic (magnetic tunneling junction \cite{MTJ_vdWM} and spin field-effect transistors \cite{spinTFET_vdWM}) as well as optoelectronic (photodiodes \cite{photodiode_vdWM} and photovoltaic detectors \cite{photovoltaic_vdWM}) applications. The  properties of vdWMs often go beyond the sum of the characteristics coming from  each individual 2-D material composing them, such as unique band alignments \cite{bandalign} \cite{bandalign2}, fast charge transport \cite{ultrafast}\cite{ultrafast2}, massive Dirac fermions \cite{Dirac}\cite{DiracHofstadter},  Hofstadter butterfly \cite{Hofstadterbutterfly} and formation of interlayer excitons \cite{exciton1} \cite{exciton2} \cite{exciton3}. Furthermore, the behaviour of vdWMs can be tuned by playing with their geometry, for example the number of layers, twisting angles between different layers, or overlap lengths \cite{2DvdW},\cite{doi:10.1021/acsnano.7b07436}.
Among the various types of vdWMs, we here consider those made of transition metal dichalcogenides (TMD).\\
All TMDs have an hexagonal crystalline structure made of three layers $X-M-X$, where X is a calchogenide and M a transition metal.   
While their electrical properties have been explored extensively, both theoretically and experimentally \cite{34d524ac883843cc8d18933b99521d10},\cite{geim2013van},\cite{frisenda2018recent},\cite{novoselov20162d},\cite{balandin2011thermal}, their thermal behavior has received much less attention although it could severely impact the applicability of these 2-D materials. For example, it has been shown that the thermal conductivity of TMDs is relatively low, about 2-3 orders of magnitude smaller than in graphene \cite{li2013thermal},\cite{Liu_2018},\cite{yan2014thermal},\cite{taube2015temperature},\cite{zhang2017thermoelectric} . Consequently, large Joule heating effects and the formation of local hot spots have been observed in TMDs \cite{pop2010energy},\cite{fu2019graphene},\cite{sahoo2013temperature}, which deteriorates the performance of devices made of such materials. Another important feature of vdWMs, is their highly anisotropic structure. The thermal conductivity of vdWMs strongly depends on the direction of propagation due to the weak van der Waals coupling that limits the heat flow along the out-of-plane axis \cite{sadeghi2016cross},\cite{pop_varshney_roy_2012}. This property might allow vdWMs to operate as directional heat spreaders, dissipating heat more efficiently along a preferred direction \cite{kim2018disorder}. 
Comparing theoretical results to experimental measurements represents a challenge for layered materials as building high-quality TMDs is a rather difficult process \cite{liu2014measurement}. This typically leads to the presence of multiple defects such as stacking errors or layer spacing variations, which induce large uncertainties in the measurement of their properties. The thermal conductivity of MoS$_2$ can vary over one order of magnitude, over several orders of magnitudes for WSe$_2$  \cite{lindroth2016thermal}. Another limitation comes from the determination of the proper thermal transport regime, often midways between the ballistic and diffusive cases. The lack of satisfactory agreement with experimental data suggests that other transport regimes might even be at play  \cite{hoogeboom2015new},\cite{barbalinardo2018thermal}. 
 In this paper, we investigate thermal transport through device configurations where the two TMD monolayers are stacked on top of each, either over the entire structure or only in their central part, as shown in Fig.~\ref{fig:vdWHs}. The electronic properties of such components have been investigated both theoretically \cite{Aron} and experimentally \cite{yan2017tunable},\cite{roy10acs}: it has been shown that they can be used as active region of band-to-band tunneling field-effect transistors with potentially a steep subthreshold swing and a high ON-state current. This aspect is therefore ignored in the present study.\\

By applying a temperature gradient between both extremities of the structure with total (TO) and partial (PO) overlap in Fig.~\ref{fig:vdWHs}, phonons start propagating from the hot to the cold side. To model this phenomena, we restrict ourselves to ballistic thermal quantum transport (QT) simulations through  vdWMs.
The required dynamical matrices are first computed via density functional perturbation theory (DFPT)
and then passed to a QT solver relying on the Wave Function (WF) or Non-equilibrium Green’s Function (NEGF) formalism to perform all thermal transport calculations. The present study intends to lay the foundations for future in-depth analyses involving more realistic effects such as anharmonic phonon decay or electron-phonon coupling.
\\The paper is organized as follows: in Section~\ref{SEC:transpModel}  an overview of the thermal transport equations is provided. Section~\ref{SEC:DFT} is dedicated to the calculation of dynamical matrices from first-principles. The influence of the inter-layer coupling is investigated in Section~\ref{sec:effect-interlayer}. Finally, in Section~\ref{SEC:results} thermal transport simulation results are presented for selected vdWMs with a partial overlap in the middle region and with total overlap. Conclusions are drawn in Section~\ref{SEC:conclusion}.


\section{Thermal Transport Model}\label{SEC:transpModel}
Thermal transport  is solved with a ballistic phonon quantum transport simulator based on the Wave Function (WF) or equivalently  Non-equilibrium Green's Function (NEGF) formalism. In the WF formalism, the equations take the form of a sparse linear system of equations $"Ax=b"$ 
\begin{equation}
    \bigg(\omega^2\mathbbm{1} - \Phi - \Pi^{RB} \bigg)\varphi = \mbox{Inj},
    \label{Phi_eq}
\end{equation}{}
which is numerically more efficient to handle than the NEGF equations in multi-dimensional structures.\\
In Eq.~(\ref{Phi_eq}), $\Pi^{RB}$ and $\mbox{Inj}$ are the retarded boundary self-energy and the Injection vector, respectively. These terms describe the phonon injection into the simulation domain through open boundary conditions. They can be determined as described in Ref.\cite{PhysRevB.74.205323}. The vector $\varphi$  contains the crystal vibrations along all Cartesian coordinates. It is of size $3Na\times N_M$, where $Na$ is the number of atoms in the system, whereas the number of columns $N_M$ indicates that $N_M$ states can be injected either from the left or the right contact at a given frequency $\omega$. Finally, $\Phi$ is the dynamical matrix of the device of size $3Na\times3Na$. A detailed description of its construction is given in Appendix \ref{Sec:DevDM-general}. All quantities $(\Pi^{RB}, \Phi, \mbox{Inj}$ and $\varphi$) depend on the phonon frequency and on the momentum $q_z$, which, for 2-D materials, is used to model the periodic, out-of-plane direction ($z$ axis in Fig.~\ref{fig:vdWHs}).
The knowledge of $\varphi$ allows to compute the frequency- and momentum-dependent transmission function $\mathcal{T}(\hbar\omega,q_z)$, from which the thermal current $I_{Th}$ can be evaluated based on the Landauer-B{\"u}ttiker formalism \cite{doi:10.1063/1.4845515},\cite{PhysRevB.79.035415}
\begin{equation}
    I_{Th} = \sum_{q_z}\frac{1}{N_{q_z}}\int \frac{d\omega}{2\pi} \mathcal{T}(\hbar\omega,q_z)\hbar\omega \bigg( b(\hbar\omega,T_L) - b(\hbar\omega,T_R)\bigg),
\end{equation}
where $N_{q_z}$ is the number of $q_z$ points used to sample the periodic direction $z$ and $b(\hbar\omega, T_{L(R)})$ is the Bose-Einstein distribution function of the left (right) contact at temperature $T_{L(R)}$.

\section{DFT Simulations}\label{SEC:DFT}
When simulating a vdWM the first step consists of constructing a suitable atomic structure, which might require assembling  two planar materials with different unit cells and/or lattice constants. Symmetry considerations play a determinant role in this regard. 
Stacking MoS$_2$ and h-BN, for example, relies on a common hexagonal cell of at least 26 atoms due to the large lattice mismatch between both materials (h-BN: 2.51 $\AA$, MoS$_2$: 3.18 $\AA$). Here, to limit the computational burden, we only investigated possible combinations of three chalcogenides (S, Se, Te) and two transition metals (Mo, W) with 2H lattice. Hence, WTe$_2$ is discarded as it is only stable in its 1T’ (metallic) form.
The combinations MoTe$_2$-MoS(Se)$_2$  are not considered either because of the large lattice mismatch between them. Combining them would require  either a unit cell with a large number of atoms or with a very high strain.
Overall, 5 homo-bilayers ( MoS$_2$-MoS$_2$, MoSe$_2$-MoSe$_2$,  WS$_2$-WS$_2$, WSe$_2$-WSe$_2$ and MoTe$_2$-MoTe$_2$) and 6  hetero-bilayers (MoS$_2$-WS$_2$, MoS$_2$-WSe$_2$, MoS$_2$-MoSe$_2$, MoSe$_2$-WS$_2$, MoSe$_2$-WSe$_2$ and WS$_2$-WSe$_2$) are analyzed. The chosen stacking order to build them is AA' (D3d group), following the same notation as in Ref.\cite{PhysRevB.89.075409}.
\begin{figure}[h!]
    \centering
    \includegraphics[width=5cm]{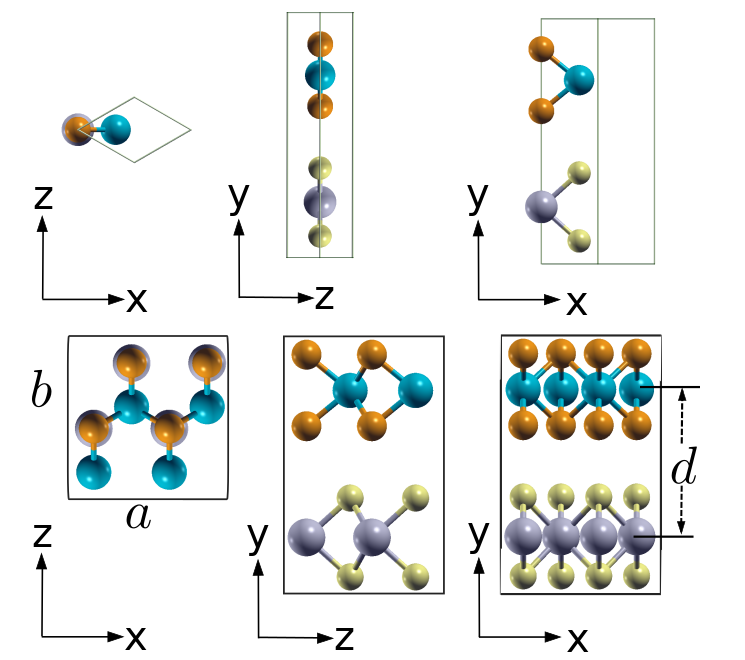}
    \caption{Bilayer hexagonal (\textit{top} panel) and orthorhombic (\textit{bottom} panel) unit cell used for transport calculation viewed under different angles. \textit{Yellow} (\textit{orange}) spheres represent the chalcogenide atoms in the \textit{bottom} (\textit{top}) layer, \textit{violet} (\textit{turquoise}) spheres the transition metal atoms in the \textit{bottom} (\textit{top}) layer.}
    \label{fig:cell_vdWH}
\end{figure}{}
The  transition metal of one layer is positioned over the chalcogenide of the other one. This mutual arrangement has been proved to be the one with lowest energy \cite{PhysRevB.89.075409},\cite{shi2016temperature},\cite{chiu10acs},\cite{ma2019first}.
Along the $c$-axis (stacking direction, aligned with the $y$-axis), a vacuum region of at least $20\,\AA$ is placed, to decouple periodic images and avoid spurious dipole interactions \cite{PhysRevB.77.245102},\cite{jiang2013fast}.
Given the extreme similarity between the TMDs under investigation, the unit cells of the bilayers is directly obtained from the unit cell of the parent materials, without the need to introduce a  supercell. Indeed, with different symmetry the lattice constant of the bilayers is found to be very close to the one of the TMDs.  
The weak inter-layer interactions between the different 2-D materials induce a very small displacement of the atomic positions with respect to the isolated TMDs. The largest displacement we calculated was along the $c$-axis and in the order of $10^{-2}\AA$. 
The hexagonal unit cell of all considered TMDs is made of 3 atoms, so that, the hexagonal unit cell of the bilayers contains 6 atoms, as illustrated in the top panel of Fig.~\ref{fig:cell_vdWH}.
Within the hexagonal cell of the bilayer, the atomic positions are accurately relaxed in order to avoid negative frequencies in the phonon spectrum. The convergence criteria was set to $10^{-8}$ \AA/eV for the force acting on each ion.  The generalized gradient approximation (GGA) of Perdew,  Burke,  and  Ernzerhof  (PBE)  \cite{PhysRevLett.77.3865} was  used as  the exchange-correlation  functional, while the  van  der  Waals  forces were  included  through  the  DFT-D2  parametrization  of  Grimme \cite{Grimme2006SemiempiricalGD}. The plane-wave cutoff  energy   was  set  to 550 eV. Subsequently, the  hexagonal cell is transformed into an orthorhombic cell, which is more convenient to perform quantum transport calculations. Such unit cells include  24 atoms, where 12  belong to one TMD and 12 to the other, as indicated in the bottom panel of Fig.~\ref{fig:cell_vdWH}. The lattice vectors and the interlayer distance of the orthorhombic cells are reported in Table \ref{tab:LatVec} for all investigated configurations.
A $2\times1\times2$ supercell discretized on a $3\times1\times3$ phonon wavevector  grid was used to compute the real space force constants with the finite displacement method.
\begin{table}[]
    \centering
    \begin{tabular}{lllc}
\hline\hline\hline
\multicolumn{4}{c}{\textbf{Lattice Vectors \& Interlayer Distance (\AA)}}\\
\hline
Bilayer&a&b&d\\
        \hline\hline
      MoS$_2$-MoS$_2$              &6.3786 & 5.5240&6.225\\
      \midrule
      WS$_2$-WS$_2$               &6.3648&5.5121&6.090\\
      \midrule
       WSe$_2$-WSe$_2$            &6.6352&5.7462& 6.502\\
      \midrule
      MoTe$_2$-MoTe$_2$           & 7.1000 &6.1488& 6.977\\
      \midrule
      MoSe$_2$-MoSe$_2$ & 6.6368 & 5.7476& 6.519\\
      \midrule
      MoS$_2$-MoSe$_2$ &6.4784 & 5.6105& 6.380 \\
      \midrule
      MoS$_2$-WS$_2$             &6.3753&5.5212& 6.153\\
      \midrule
      MoS$_2$-WSe$_2$            &6.5006&5.6297& 6.320\\
      \midrule
      MoSe$_2$-WS$_2$ &6.5008 & 5.6299& 6.307\\
      \midrule
      MoSe$_2$-WSe$_2$ &6.6360&  5.7469& 6.475 \\
      \midrule
      WS$_2$-WSe$_2$ &6.5000 & 5.6292& 6.253\\
\hline
    \end{tabular}
    \caption{Orthorhombic unit cell dimensions of the investigated vdWMs where $a$ and $b$ refer to the in-plane axis, aligned with the $x-$ and $z-$axis, respectively. The variable $d$ indicates the distance along the $y-$axis between chalcogenides belonging to different layers, as depicted in Fig.~\ref{fig:cell_vdWH}  }
    \label{tab:LatVec}
\end{table}

\begin{figure}[h!]
    \centering
    \includegraphics[width=\linewidth]{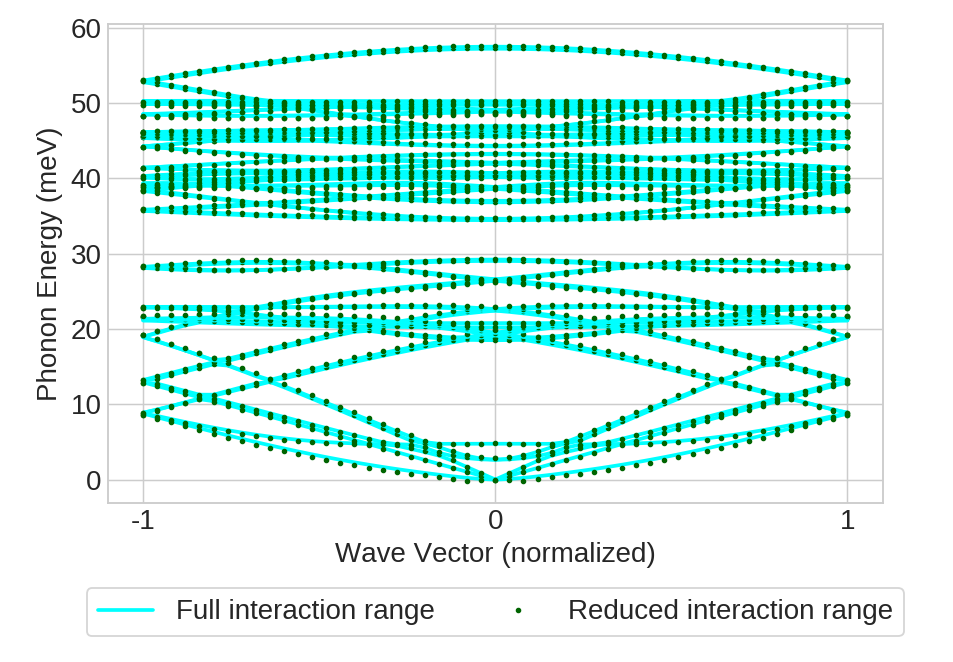}
    \caption{Phonon dispersion of a MoS$_2$-MoS$_2$ homo-bilayer  computed with PHONOPY \cite{phonopy} keeping the full inter-atomic interaction range (\textit{cyan} lines) or truncating it beyond $r_{cut}=6.8 \AA$ (\textit{green} dots).}
    \label{fig:cutoff}
\end{figure}

All DFT calculations were carried out with the Vienna \textit{Ab initio} Software Package (VASP) \cite{PRB_54_11169},\cite{KRESSE199615} and PHONOPY \cite{phonopy}. All inter-atomic interactions were cut off beyond a radius $r_{cut}=6.8\AA$ in the simulated vdWMs. It has been verified that this truncation does not affect the phonon bandstructure by more than 10\%, ensuring accurate quantum transport calculations. As an example the dispersion of  MoS$_2$-MoS$_2$ homo-bilayer is reported in Fig.~\ref{fig:cutoff} with the full and truncated interaction range. An excellent agreement can be observed between both sets of curves.
{}


\section{Effect of the Inter-Layer Coupling}\label{sec:effect-interlayer}
Next, the impact of the inter-layer coupling (van der Waals forces) on the Interatomic Force Constants and phonon properties is studied.
\begin{figure}[h!]
 \centering\includegraphics[width=\linewidth]{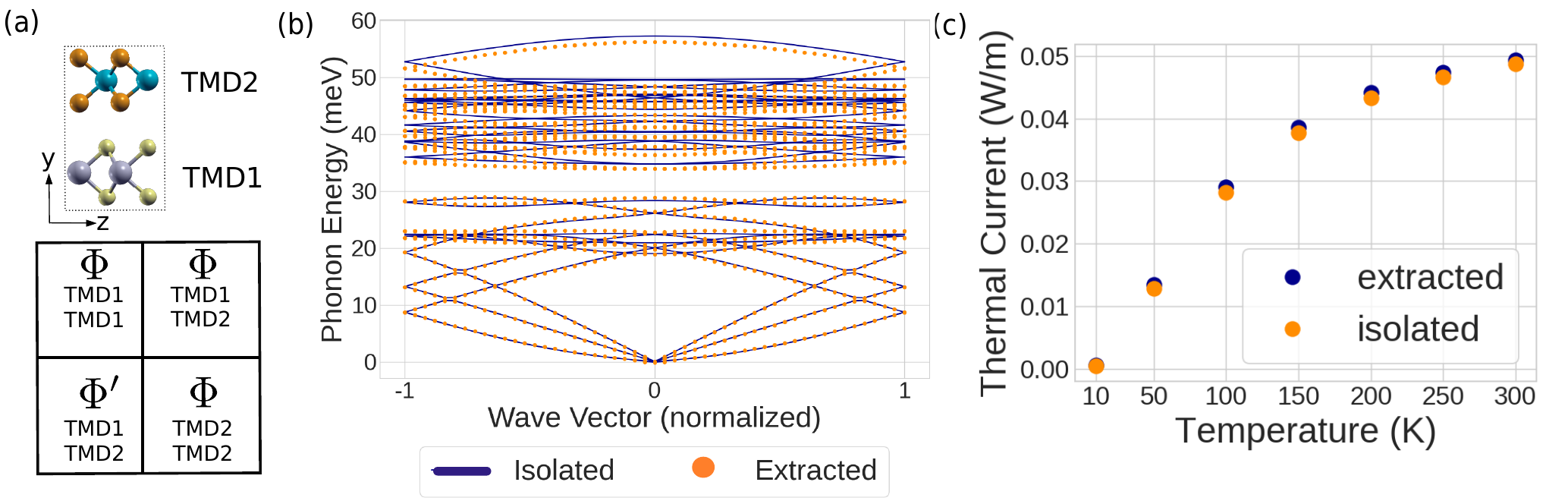}
    \caption{(a) Bilayer structure made of TMD1 and TMD2 and corresponding schematics of the Interatomic Force Constants. (b) Phonon dispersion of monolayer MoS$_2$ computed from an isolated structure (\textit{blue} lines) and when the dynamical matrix is extracted from MoS$_2$-WS$_2$ hetero-bilayer (\textit{orange} dots). (c) Ballistic thermal current flowing through a MoS$_2$ monolayer calculated with the phonon dispersions from (b) as a function of the temperature. A thermal gradient $\Delta T = 0.1\,K$ is applied between both extremities of the structure.}
    \label{fig:MoS2extracted_I}
\end{figure}
In Fig.~\ref{fig:MoS2extracted_I}(a) the schematics a vdWM bilayer is shown together with the corresponding dynamical matrix. The diagonal blocks contain the intra-layer interactions and the off diagonal blocks  the inter-layer coupling elements. The average magnitude of the intra-layer blocks ($\Phi_{\mbox{TMD1-TMD1}}$ and $\Phi_{\mbox{TMD2-TMD2}}$) entries exceeds that of the inter-layer blocks by a factor 500. Motivated by this observation we examined whether the properties of each individual layer could be retrieved from those of the stack. In other words, we wanted to verify whether the intra-layer matrices  $\Phi_{\mbox{TMD1-TMD1}}$ ($\Phi_{\mbox{TMD2-TMD2}}$) in Fig.~\ref{fig:MoS2extracted_I}(a) could accurately describe the phonon dispersion and the ballistic thermal current flowing through a monolayer of TMD1(TMD2). 
To compute the ballistic thermal current a gradient $\Delta T = 0.1 K$ was applied between the temperature of the left ($T_L$) and the right ($T_R$) contacts, while varying $T_L$ between 10 and 300 $K$. The phonon transport equations were solved by discretizing the periodic direction $z$ with $N_{qz} = 51$ momentum points and  by considering $N_M = 150$ frequency points.
As an example, a MoS$_2$-WS$_2$ vdWM was chosen. The phonon bandstructure and thermal current of a MoS$_2$ monolayer were extracted and compared to data directly obtained from a calculation involving individual, isolated layer calculation. The results are reported in Fig.~\ref{fig:MoS2extracted_I}(b) and Fig.~\ref{fig:MoS2extracted_I}(c), respectively. The same type experiment was repeated for other 2-D material combinations: in all cases, the error between the data derived from the vdWM dynamical matrix and from a true monolayer remains below 10\%.
We continued our analysis with the thermal current flowing through vdWM bilayers overlapping over their entire surface, as in Fig.~\ref{fig:vdWHs}(a), and through the individual 2-D monolayer constituting them. For brevity, such bilayer structures with fully overlapping monolayers are referred to TO, while the isolated layers are labeled  1 and 2, where 1(2) indicate the first(second) TMD in the vdWM. The resulting bilayer is of the form  $M^{(1)}X^{(1)}_2-M^{(2)}X^{(2)}_2$. 
All calculations were performed at $300\,K$ with $1\,K$ temperature difference between the left and right contact. Hence, $I_{Th,TO}$ is the current flowing through the TO structure, and $I_{Th,1(2)}$ through first(second) monolayers. We found that the relative error between $I_{Th,1}+I_{Th,2}$ and  $I_{Th,TO}$ does not exceed 10\%, as can be seen in Table \ref{tab:TO}.

\begin{table}[]
    \centering
    \begin{tabular}{llllrr}
    \hline\hline\hline
\multicolumn{5}{c}{\textbf{Thermal Current [W/m]}}\\
\hline
Bilayer &$I_{Th,1}$ &$I_{Th,2}$& $I_{Th,TO}$ &$I_{Th,1}$& Error\\
&&&&$+I_{Th,2}$&[\%]\\
        \hline\hline
MoS$_2$-MoS$_2$  &0.50037&0.50037&0.9855 &1.0007 &2\\\hline
WS$_2$-WS$_2$ &0.3986 & 0.3986 & 0.7771  & 0.7972&3\\\hline
WSe$_2$-WSe$_2$ & 0.3166 & 0.3166 & 0.6145  & 0.6332&2 \\\hline
MoTe$_2$-MoTe$_2$ & 0.2567 & 0.2567 & 0.4839 & 0.5134&6\\\hline
MoSe$_2$-MoSe$_2$  & 0.3472 & 0.3472 & 0.6676 & 0.6944&4 \\\hline\hline
MoS$_2$-MoSe$_2$ & 0.4994 & 0.3802 & 0.863  & 0.8796&2 \\\hline
MoS$_2$-WS$_2$  & 0.4982 & 0.3762 & 0.8222 & 0.8744&6\\ \hline
MoS$_2$-WSe$_2$ & 0.4812 & 0.3275 & 0.7417  & 0.8087&8 \\ \hline
MoSe$_2$-WS$_2$  & 0.3789 & 0.3821 & 0.7393 & 0.7610&3 \\\hline
MoSe$_2$-WSe$_2$  & 0.3681 & 0.3132 & 0.6589  & 0.6813&3 \\\hline
WS$_2$-WSe$_2$  & 0.3817 & 0.3359 & 0.6924  & 0.7176 &4\\\hline   \end{tabular}
    \caption{Comparison between the thermal current flowing through  2-D individual monolayers ($I_{Th,1(2)}$) and through vdWM bilayers overlapping over the entire surface ($I_{Th,TO}$). The last column reports the relative error between $I_{Th,1}+I_{Th,2}$ and  $I_{Th,TO}$}
    \label{tab:TO}
\end{table}


\section{Transport Simulations and Results}\label{SEC:results}

In this Section, the ballistic thermal transport properties of vdWMs with partial overlap (as in Fig.~\ref{fig:vdWHs}(b) ) are reported and analysed. The developed methodology allows to simulate such device configurations that are more complex and potentially more interesting than TO channels , where the two  TMD monolayers are stacked on top of each other over their entire surface.
In geometries with a partial TMD overlap, three distinct regions can be identified: one with an isolated monolayer TMD1, on the left, one with an isolated monolayer TMD2 on the right and one with a bilayer vdWM TMD1+TMD2 in the middle, as sketched in Fig.~\ref{fig:vdWHs}(b). In fact, the structure in Fig.~\ref{fig:vdWHs}(b) is obtained by removing atoms from the bilayer in Fig.~\ref{fig:vdWHs}(a) and by eliminating the corresponding entries from the dynamical matrix. A detailed description of the construction procedure and of the assembling of its device dynamical matrix is given in Appendix \ref{Sec:inhomog-DDM}. 
While it was demonstrated in the previous Section that the interlayer coupling has a limited influence on the thermal current flowing through a vdWM with a TO structure, the situation is radically different in the case of a partial overlap, even if the two layers are identical. In geometries similar to the one shown in Fig.~\ref{fig:vdWHs}(b), without inter-layer coupling, there would be no thermal current, thus making the van der Waals interactions a critical component of the heat flow.
At the microscopic level, the different configuration of the three device regions forces the phonons to be transferred from one layer to the other. 
This does not prevent the application of our approach as  the inter-layer coupling is fully captured in the overlap region, which ensures an accurate description of thermal transport there. 
As will be shown in the following, the thermal transport properties of a bilayer channel with a partial overlap  cannot be predicted from the phonon bandstructure of the individual monolayers that compose it, contrary to structures where the TMDs are stacked over their entire surface. Atomistic quantum transport simulations are required to reveal their behaviour. 

\begin{figure}[h!]
    \centering
    \includegraphics[width=0.5\textwidth]{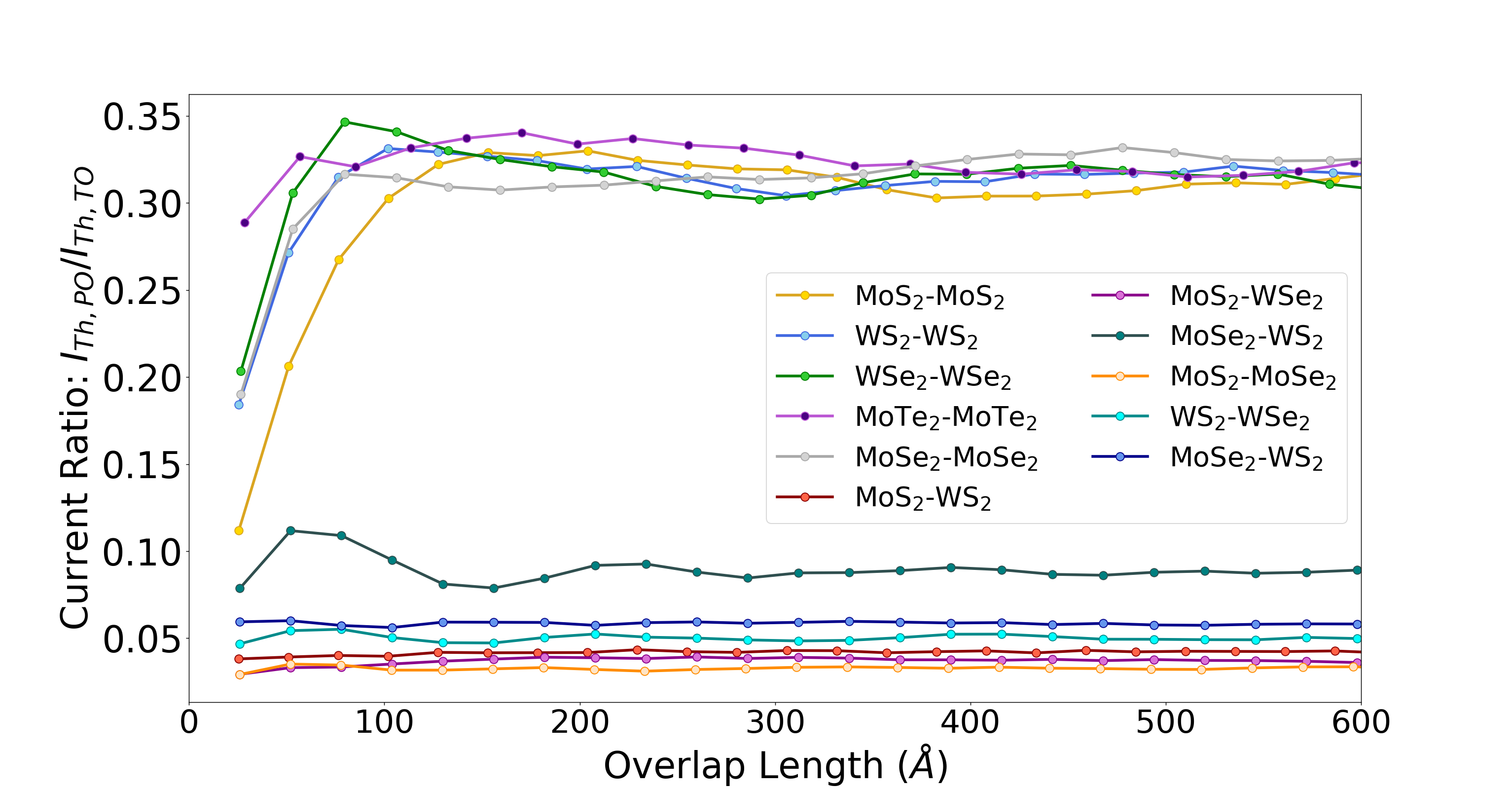}
    \caption{ Ballistic thermal current ratio $I_{Th,PO}/I_{Th,TO}$ for all the homo- and hetero-bilayer vdWMs considered in this work.}
    \label{fig:current_vs_overlap}
\end{figure}

As starting point,  the impact of the overlap length on the ballistic current is presented in  Fig.~\ref{fig:current_vs_overlap}, the goal being to study how this parameter affects the current magnitude and whether representative features can be identified. A slight dependence of the current on the overlap length can be observed in some of the homo-bilayers (both TMDs are identical). Below 100-150 $\AA$, the thermal current linearly increase with the overlap length before saturating. This seems to indicate that a minimum overlap length is needed so that all phonon modes can be transferred from one layer to the other. Note that very short overlap lengths are probably very challenging to achieve experimentally
 \cite{weiss2016van} and therefore practically of limited relevance.  Hence, from now on, we restrict our  investigations to devices with an overlap length of approximately 300 $\AA$. With such dimensions, for all bilayers the simulation results are independent from the length of the overlap region. For brevity, we refer to bilayer structures with 300 $\AA$ of overlap as PO.
 
Table~\ref{tab:PO} reports three relevant quantities concerning the current flowing through these structures: (i) the ratio between $I_{th,PO}$ and $I_{th,TO}$, (ii) that between $I_{th,PO}$ and $I_{th,1}$ and (iii) that between $I_{th,PO}$ and $I_{th,2}$. Together with Fig.~\ref{fig:current_vs_overlap}, these results highlight that the ballistic thermal current flowing through a PO channel is reduced by about 70\% as compared to TO channel in case of homo-bilayers
(MoS$_2$-MoS$_2$, MoSe$_2$-MoSe$_2$, WS$_2$-WS$_2$, WSe$_2$-WSe$_2$,  MoTe$_2$-MoTe$_2$). A more significant decrease of approximately 95\% is observed for hetero-bilayers (MoS$_2$-WS$_2$, MoS$_2$-WSe$_2$, MoS$_2$-MoSe$_2$, MoSe$_2$-WS$_2$, MoSe$_2$-WSe$_2$, WS$_2$-WSe$_2$). Peraphs even more important, the thermal current through PO strucutures is smaller than that flowing through either of the monolayers constituting it, by 40\% for the homo-bilayers, much more for the hetero-bilayers.
{}

\begin{figure}[h!]
    \centering
    \includegraphics[width=0.5\textwidth]{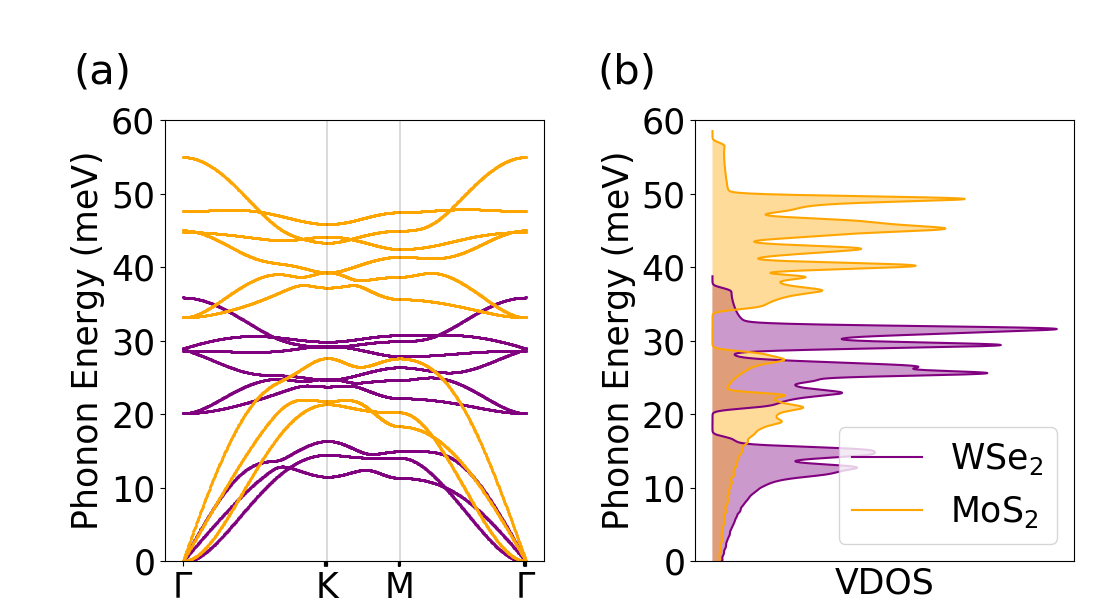}
    \caption{(a) Phonon dispersion of MoS$_2$ (\textit{orange}) and WSe$_2$ (\textit{purple}) monolayers. (b) Corresponding Vibrational Density-of-States (VDOS).}
    \label{fig:VDOS}
\end{figure}

\begin{table}[h!]
    \centering
    \begin{tabular}{lccc}
\hline\hline\hline
\multicolumn{4}{c}{\textbf{Thermal Current}}\\
\hline
Bilayer&  $I_{Th,PO}$/$I_{Th,TO}$ &$I_{Th,PO}$/$I_{Th,1}$ &$I_{Th,PO}$/$I_{Th,2}$\\
        \hline\hline
MoS$_2$-MoS$_2$& 0.3147&0.6199&0.6199\\\hline
WS$_2$-WS$_2$  & 0.3070 & 0.5986 & 0.5986\\\hline
WSe$_2$-WSe$_2$ & 0.3115 & 0.6045 & 0.6045 \\\hline
MoTe$_2$-MoTe$_2$  &  0.3224 & 0.6077 & 0.6077\\\hline
MoSe$_2$-MoSe$_2$  & 0.3168 & 0.6092 & 0.6092\\\hline\hline 
MoS$_2$-MoSe$_2$  & 0.0335 & 0.0579 & 0.0760\\\hline
MoS$_2$-WS$_2$    & 0.0429 & 0.0708 & 0.0937\\ \hline
MoS$_2$-WSe$_2$  & 0.0386 & 0.0595 & 0.0874\\ \hline
MoSe$_2$-WS$_2$ & 0.0877 & 0.1710 & 0.1696\\\hline
MoSe$_2$-WSe$_2$ & 0.0596 & 0.1068 & 0.1255\\\hline
WS$_2$-WSe$_2$  & 0.0487 & 0.0883 & 0.1003\\\hline
\hline
    \end{tabular}
    \caption{Ratio between the ballistic thermal current flowing through the PO ($I_{th,PO}$) and TO ($I_{th,TO}$) structures and through the PO vdWMs and the isolated monolayers ($I_{th,1(2)}$) they are made of.}
    \label{tab:PO}
\end{table}

The extremely high thermal current reduction in case of hetero-bilayers can be attributed to the fact that in the ballistic limit of transport only states existing throughout the whole structure can be transmitted from one contact to the other~\cite{ballistic}. This can be qualitatively explained by inspecting the phonon bandstructure and the vibrational density-of-states (VDOS) of the two TMDs composing the vdWM. As an example, in Fig.~\ref{fig:VDOS}, the data for a MoS$_2$-WSe$_2$ hetero-bilayer is presented. While MoS$_2$ exhibits a mini-gap around the phonon energy $\hbar \omega = 30 \,meV$ ($ 27 < \hbar \omega < 32 \,meV$) and its energy range goes up to $\hbar\omega_{MAX}\,=\,57\,meV$, it can be seen that the energy window without VDOS for WSe$_2$ is different, i.e. $ 17 < \hbar \omega < 20 \,meV$, as well as its maximum energy ($\hbar\omega_{MAX}\,=\,38\,meV$). As a consequence, a phonon with energy $\hbar \omega = 45 \,meV$ injected from MoS$_2$ will not be able to be transferred to the WSe$_2$ monolayer, which reduces the number of states available for transport and the resulting thermal current. Further comparisons similar to Fig.~\ref{fig:VDOS} are provided in the Supplementary Materials, for other 2-D material combinations. 
\begin{figure}
    \centering
    \includegraphics[width=\linewidth]{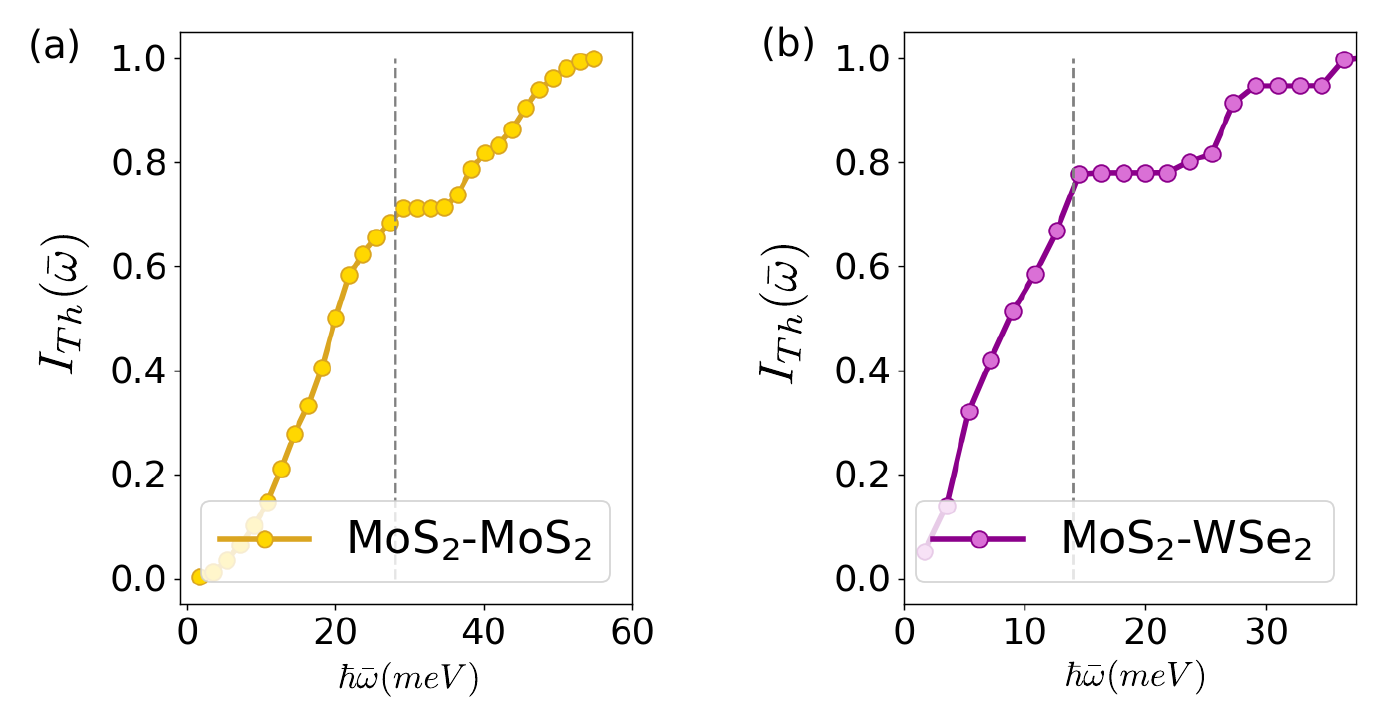}
    \caption{(a) Cumulative ballistic thermal current of the homo-bilayer MoS$_2$-MoS$_2$ PO structure computed up the phonon energy $\hbar \Bar{\omega}$, as described in Eq.~(\ref{cumcurr}). The dashed line indicates the end of the acoustic phonon branches of the TMD with the lower energy range. (b) Same as (a) but for the hetero-bilayer MoS$_2$-WSe$_2$ PO structures}
    \label{fig:CumCurr}
\end{figure}

To continue our analysis, we examined on the energy (frequency) dependence of the thermal current. Fig.~\ref{fig:CumCurr} shows the values of the cumulative ballistic thermal current $I_{Th}(\Bar{\omega})$ of the homo-bilayer  MoS$_2$-MoS$_2$ and the  hetero-bilayer MoS$_2$-WSe$_2$  computed according to the following equation
\begin{equation}
\begin{split}
    &I_{Th}(\Bar{\omega}) = \\ &\frac{\sum_{q_z}\int_{0}^{\hbar \Bar{\omega}} \frac{d\omega}{2\pi} \mathcal{T}(\hbar\omega,q_z)\hbar\omega \bigg( b(\hbar\omega,T_L) - b(\hbar\omega,T_R)\bigg) }{\sum_{q_z}\int_{0}^{\hbar \omega_{MAX}} \frac{d\omega}{2\pi} \mathcal{T}(\hbar\omega,q_z)\hbar\omega \bigg( b(\hbar\omega,T_L) - b(\hbar\omega,T_R)\bigg)}.
    \label{cumcurr}
    \end{split}
\end{equation}
 In Eq.~(\ref{cumcurr}) $\hbar \omega_{MAX}$ is the highest phonon energy of the vdWM. From this analysis it clearly appears that the major contribution to the ballistic thermal current comes from the acoustic branches.  Similar analyses of all considered vdWMs with a PO configuration can be found in the Supplementary Materials. 
 
\begin{figure}[h!]
    \centering
    \includegraphics[width=0.5\textwidth]{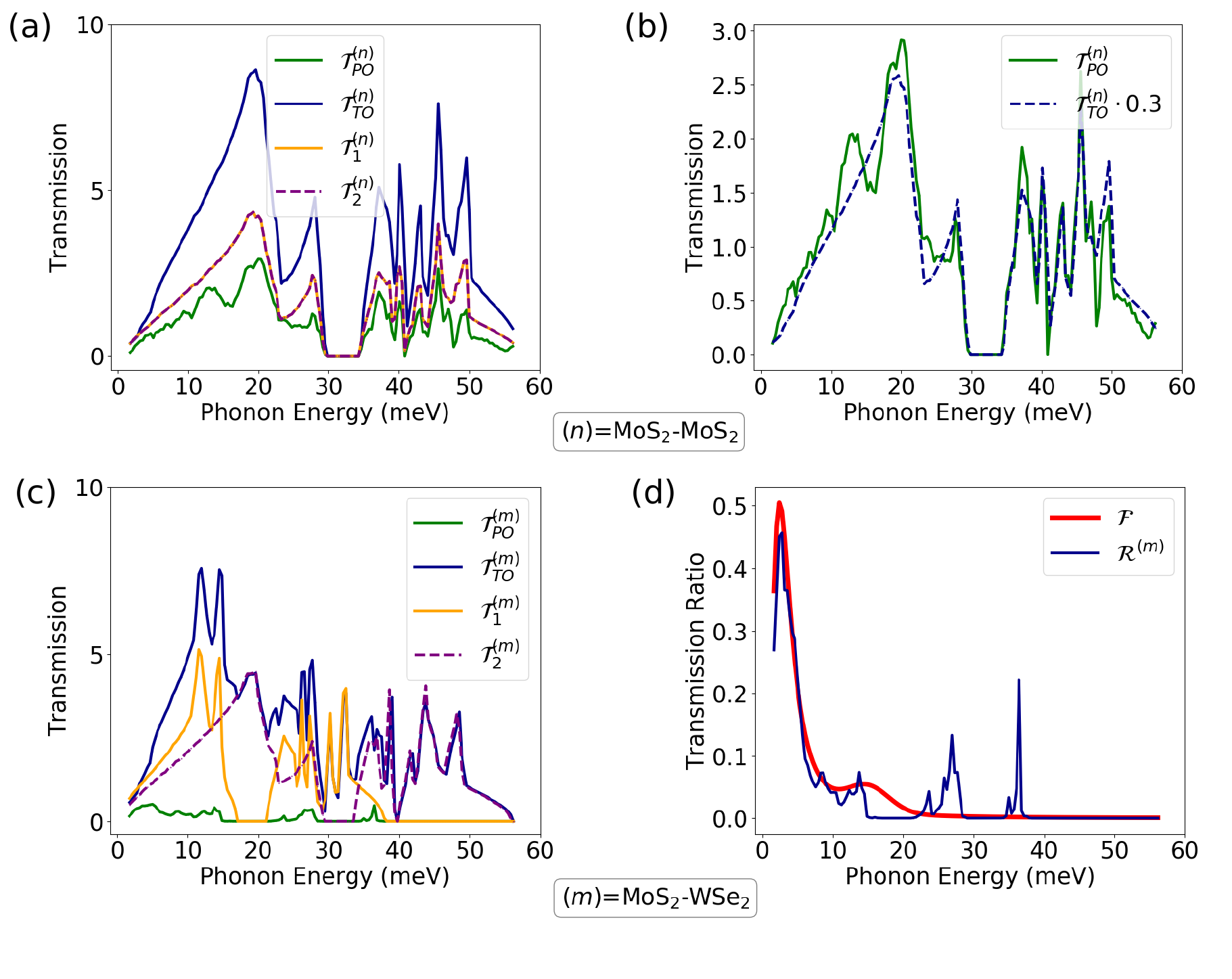}
    \caption{(a)  Transmission function $\mathcal{T}^{(n)}_{C}(\hbar \omega)$ of the homo-bilayer vdWM with $(n)$ =MoS$_2$-MoS$_2$ case with different channel configuration $(C)=$ PO, TO, 1 and 2. (b) Comparison of $\mathcal{T}^{(n)}_{PO}(\hbar \omega)$ and $\mathcal{T}^{(n)}_{TO}(\hbar \omega)$ scaled by a factor 0.3, for the homo-bilayer $(n)=$ MoS$_2$-MoS$_2$ vdWM. (c) Same as (a), but for the hetero-bilayer vdWM with $(m)$=MoS$_2$-WSe$_2$ structure.(d) Transmission function ratio  $\mathcal{R}^{(m)}(\hbar \omega)$= $\mathcal{T}^{(m)}_{PO}(\hbar \omega)/\mathcal{T}^{(m)}_{TO}(\hbar \omega)$ for the hetero-bilayer $(m)=$ MoS$_2$-WSe$_2$ (\textit{blue} line) and  the function $\mathcal{F}$ described by Eq.~(\ref{fit}) (\textit{red} line) }
    \label{fig:Ratio}
\end{figure}

As final step, we studied the energy distribution of the total transmission function $\mathcal{T}(\hbar \omega) = \frac{1}{N_{q_z}}\sum_{q_z}\mathcal{T}(\hbar \omega, q_z)$.
To distinguish the different cases under investigation, we introduced the following notation: be $(n)\in$ \{MoS$_2$-MoS$_2$, MoSe$_2$-MoSe$_2$, WS$_2$-WS$_2$, WSe$_2$-WSe$_2$ and  MoTe$_2$-MoTe$_2$\} the index referring to the homo-bilayers, $(m)\in$ \{MoS$_2$-WS$_2$, MoS$_2$-WSe$_2$, MoS$_2$-MoSe$_2$, MoSe$_2$-WS$_2$, MoSe$_2$-WSe$_2$ and WS$_2$-WSe$_2$\} the index referring to the hetero-bilayers and $(C)\in$ \{PO, TO, 1, 2\} the index indicating the different channel configurations (partial overlap, total overlap, only first TMD, only second one). 
Fig.~\ref{fig:Ratio}(a) reports the energy-resolved transmission functions $\mathcal{T}^{(n)_{(C)}}(\hbar \omega)$ of the homo-bilayer $(n)=$ MoS$_2$-MoS$_2$ with all possible channel configurations. Note that channels 1 and 2 are identical in this case. 
In Fig.~\ref{fig:Ratio}(b), the comparison between $\mathcal{T}^{(n)_{(PO)}}(\hbar \omega)$ and $\mathcal{T}^{(n)_{(TO)}}(\hbar \omega)$ scaled by a factor 0.3, indicates that the energy profile of the transmission function in the PO channel  can be obtained, as first approximation, by scaling the transmission function of the TO channel by a constant, energy-independent factor.

This behaviour can be interpreted by recalling the fact that in an homo-bilayer with a PO configuration, the initial and final states are the same. As they are mixed in the overlap region each state is degenerate.
It should be emphasized that the scaling factor, 0.3, is in agreement with results of Table~\ref{tab:PO} and remains constants regardless of the TMD monolayers composing the homo-vdWMs (see the Supplementary Materials).
 Fig.~\ref{fig:Ratio}(c) presents the same analysis as Fig.~\ref{fig:Ratio}(a), but for the hetero-bilayer $(m)=$ MoS$_2$-WSe$_2$ structure, where channels 1 and 2 differ. Contrary to the homo-bilayer case, the energy-resolved transmission function of PO configuration, $\mathcal{T}^{(m)_{(PO)}}(\hbar \omega)$ cannot be obtained by scaling $\mathcal{T}^{(m)_{(TO)}}(\hbar \omega)$ or $\mathcal{T}^{(m)_{(1)}}(\hbar \omega)$ and $\mathcal{T}^{(m)_{(2)}}(\hbar \omega)$. Indeed, a combination of the transmission functions of the isolated monolayers would only account for the presence of mini-gaps as well as for the extension of the energy spectra already noticed in Fig.~\ref{fig:VDOS}. By building the transmission function ratio 
\begin{equation}
    \mathcal{R}^{(m)}(\hbar \omega) = \frac{\mathcal{T}^{(m)_{(PO)}}(\hbar \omega)}{\mathcal{T}^{(m)_{(TO)}}(\hbar \omega)},
\end{equation}
which is reported in Fig.~\ref{fig:Ratio}(d) for  MoS$_2$-WSe$_2$, a peak can be detected at $5\,meV$, followed by bumps of much smaller intensity. We repeated the same calculation for all hetero-bilayers, as reported in the Supplementary Materials, we found that all $\mathcal{R}^{(m)}(\hbar \omega)$ exhibit a peak with a magnitude of about 0.5, at approximately $\hbar\omega=5\,meV$, followed by bumps, whose behaviour depends on the material combination. Based on this observation we computed the average transmission function ration
\begin{equation}\label{ravg}
    \mathcal{R}_{AVG}(\hbar \omega) = \frac{1}{N_m}\sum_m \mathcal{R}^{(m)}(\hbar \omega)
\end{equation}
where $N_m\,=\,6$ is the number of hetero-bilayers that we investigated. $\mathcal{R}_{AVG}(\hbar \omega)$ can be fitted using the expression:
\begin{equation}\label{fit}
    \mathcal{F}(\hbar\omega) = \mathcal{L}\big(\hbar \omega, \mu_1, \eta, A\big) + \frac{1}{N}e^{-(\hbar \omega-\mu_2)/2\sigma^2},
\end{equation}
where $\mathcal{L}$ is the Landau distribution function, as defined in Ref.~\cite{pylandau}, using the following parameters ($\mu_1, \eta, A, \mu_2, \sigma, N$)= (1749.97, 549.68, 0.505, 15, 3, 0.04). The function $ \mathcal{F}(\hbar\omega)$ is plotted  in Fig.~\ref{fig:Ratio}(d).
As ${\mathcal{F}}$ approximates the ratio between the $\mathcal{T}^{(m)}_{PO}(\hbar\omega)$ and $\mathcal{T}^{(m)}_{TO}(\hbar\omega)$, from the knowledge of $\mathcal{T}^{(m)}_{TO}(\hbar\omega)$, which can be derived from $\mathcal{T}^{(m)}_{1}(\hbar\omega)$ and $\mathcal{T}^{(m)}_{2}(\hbar\omega)$, we can estimate the thermal current flowing through a hetero-bilayer with a PO configuration
\begin{equation}\label{If}
\begin{split}
    I^{(m)}_{\mathcal{F}} = \int_{0}^{\hbar\omega_{MAX}} \frac{d\omega}{2\pi}& \mathcal{T}^{(m)}_{TO}(\hbar\omega)\mathcal{F}(\hbar\omega)\hbar\omega \cdot \\
    &\cdot\bigg( b(\hbar\omega,T_L) - b(\hbar\omega,T_R)\bigg).
\end{split}
\end{equation}
The ratio between  $I^{(m)}_{\mathcal{F}}$ and the corresponding ballistic thermal current  $I^{(m)}_{PO}$ obtained with a quantum transport simulation is listed in Table~\ref{tab:TOmod}. 
It can be observed that the non-uniformly scaled transmission function $\mathcal{T}^{(m)}_{TO}(\hbar\omega)\mathcal{F}(\hbar\omega)$ of Eq.~(\ref{If}) can reproduce the quantum mechanical results with an error that does not exceed 35\%. Only MoSe$_2$-WS$_2$ does not follow this trend, due to a large transmission contribution  in the energy region 10 $\leq \hbar\omega \leq 20\,meV$ which is not correctly captured by the function $\mathcal{F}(\hbar\omega)$. Still, it can be deduced that the hetero-vdWMs with a PO configuration mainly act as low-pass filter of the corresponding TO transmission function.
\begin{table}[h!]
    \centering
    \begin{tabular}{ll}
\hline\hline\hline
\multicolumn{2}{c}{\textbf{Thermal Current}}\\
\hline
Hetero-Bilayer $(m)$&	$I^{(m)}_{\mathcal{F}}/I^{(m)}_{PO}$\\\hline\hline
MoS$_2$-WS$_2$    &0.65\\\hline
MoS$_2$-WSe$_2$   &0.83\\\hline
MoSe$_2$-WS$_2$  &0.37\\\hline
MoS$_2$-MoSe$_2$  &0.82\\\hline
WS$_2$-WSe$_2$   &0.73\\\hline
MoSe2-WSe$_2$  &0.63\\\hline
\hline
    \end{tabular}
    \caption{Ratio between $I^{(m)}_{\mathcal{F}}$ and the corresponding ballistic thermal current $I^{(m)}_{PO}$ calculated with a quantum transport simulation for hetero-bilayer vdWMs with a PO configuration. }
    \label{tab:TOmod}
\end{table}

\begin{figure}[h!]
    \centering
    \includegraphics[width=0.5\textwidth]{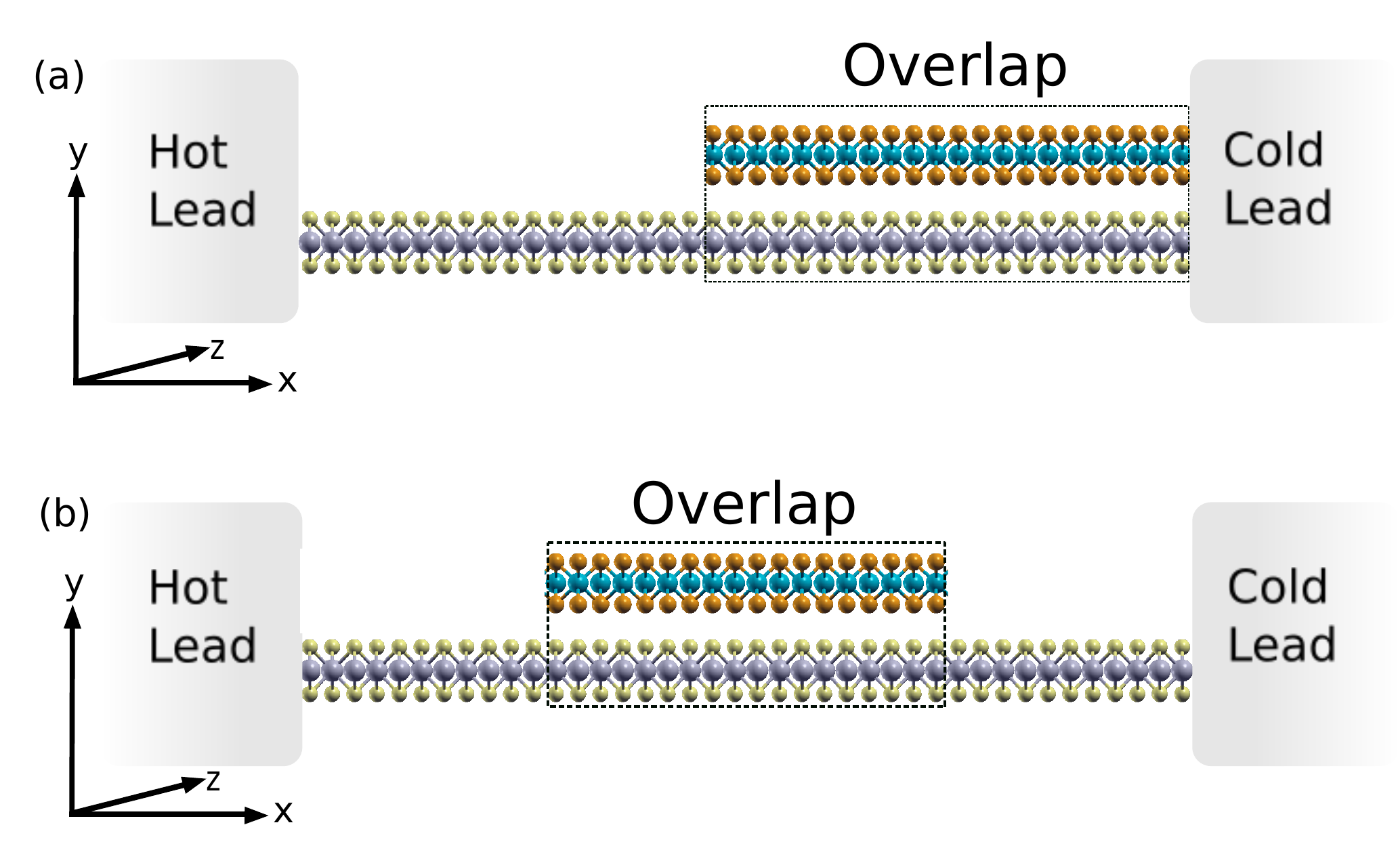}
    \caption{(a) vdWM made of two TMDs, where one TMD extends over the whole device length, while the other goes from one reservoir to the middle of the device. Such structure is referred to as 1O(2O) if TMD1(TMD2) is connected to both reservoirs. (b) vdWM made of two TMDs, one covering the entire device length and the other only existing in the central region, without connection to the reservoirs. Such structures are labeled 1O1(2O2) if TMD1(TMD2) is extended over the whole length}
    \label{fig:other}
\end{figure}

\begin{table}[h!]
    \centering
\begin{tabular}{lcc}
    \hline\hline\hline
\multicolumn{3}{c}{\textbf{Thermal Current [W/m]}}\\
\hline
Bilayers &		$I_{Th,1O}/I_{Th,1}$ &	$I_{Th,2O}/I_{Th,2}$ \\
        \hline\hline
    MoS$_2$-MoS$_2$ &	0.95 &	0.95\\\hline
 	WS$_2$-WS$_2$ &	0.92 &	0.93\\\hline
 	WSe$_2$-WSe$_2$ &	0.92 &	0.92\\\hline
 	MoTe$_2$-MoTe$_2$ &	0.90 &	0.90\\\hline
 	MoSe$_2$-MoSe$_2$ &	0.97&	0.97\\\hline\hline
 	MoS$_2$-MoSe$_2$ &	0.93 &	0.94\\\hline
 	MoS$_2$-WS$_2$ &	0.92&	0.91\\\hline
 	MoS$_2$-WSe$_2$ &	0.95 &	0.93\\\hline
 	MoSe$_2$-WS$_2$ &	0.93 &	0.90\\\hline
 	MoSe$_2$-WSe$_2$ &	0.94 &	0.89\\\hline
  	WS$_2$-WSe$_2$ 	&	0.88 &	0.90\\\hline

\hline
\end{tabular} 
    \caption{Ratio between the ballistic thermal current flowing through the 1O ($I_{th,1O}$) structure and TMD1 ($I_{th,1}$) and through the 2O ($I_{th,2O}$) ones and TMD2 ($I_{th,2}$)}
    \label{tab:others2}
\end{table}

\begin{table}[h!]
    \centering
\begin{tabular}{lcc}
    \hline\hline\hline
\multicolumn{3}{c}{\textbf{Thermal Current [W/m]}}\\
\hline
Bilayers &	$I_{Th,1O1}/I_{Th,1}$& 	$I_{Th,2O2}/I_{Th,2}$  \\
        \hline\hline
MoS$_2$-MoS$_2$ &	0.57&	0.57	\\\hline
 	WS$_2$-WS$_2$ &	0.58 &	0.58 \\\hline
 	WSe$_2$-WSe$_2$ &	0.59 &	0.59 \\\hline
 	MoTe$_2$-MoTe$_2$&	0.59 &	0.59 \\\hline
 	MoSe$_2$-MoSe$_2$ &	0.58 &	0.58 \\\hline\hline
 	MoS$_2$-MoSe$_2$ &	0.90 &	0.91 \\\hline
 	MoS$_2$-WS$_2$ &	0.90 &	0.93 \\\hline
 	MoS$_2$-WSe$_2$ &	0.92 &	0.88 \\\hline
 	MoSe$_2$-WS$_2$ &	0.86 &	0.83 \\\hline
 	MoSe$_2$-WSe$_2$ &	0.86 &	0.85 \\\hline
 	WS$_2$-WSe$_2$ 	& 0.83 &	0.86 \\\hline
\hline
\end{tabular} 
   \caption{Ratio between the ballistic thermal current flowing through the 1O1 structure ($I_{th,1O1}$)  and TMD1 ($I_{th,1}$) and through the 2O2 structure ($I_{th,2O2}$) and TMD2 ($I_{th,2}$).}
    \label{tab:others3}
\end{table}

Besides the PO structure analyzed in this work, two other arrangements of vdWMs can be envisioned. The first possibility is illustrated in Fig.~\ref{fig:other}(a), where one TMD is extended along the whole device length, while the other covers only a part of the distance separating both electrodes. Such structures are referred to as 1O (2O) channels, if the TMD extended along the whole length is TMD1 (TMD2). In the second possibility, depicted in Fig.~\ref{fig:other}(b), one TMD extends along the whole structure, whereas the second one is only present in the central region and not connected to any lead. This configuration is labeled 1O1 (2O2) channels if the TMD extended along the whole length is TMD1 (TMD2). The ratio between the ballistic thermal current flowing  through the 1O ($I_{th,1O}$) structure and TMD1 ($I_{th,1}$) and through the 2O ($I_{th,2O}$) structure and TMD2 ($I_{th,2}$) are presented in Table~\ref{tab:others2}. The latter shows that 1O (2O) devices behave as the TMD extended over the whole length.
Table~\ref{tab:others3} reports the ratio between the ballistic thermal current flowing through the 1O1 structure ($I_{th,1O1}$) and TMD1 ($I_{th,1}$) as
well as through the 2O2 structure ($I_{th,2O2}$) and TMD2 ($I_{th,2}$). In the homo-bilayer cases, both the 1O1 and 2O2 configurations
allow for a similar percentage of the thermal current to flow through them (with respect to $I_{th,1}$ and $I_{th,2}$ ) as the comparable
PO structures, about 60\%. The situation is different in the hetero-bilayer devices where the thermal current is found to reach
more than 80\% of the value corresponding to the TMD connected to both contacts. The strength of the inter-layer coupling
explains this behavior. Because these interactions are larger in homo-bilayer structures, the impact of stacking a
disconnected TMD layer on top of a connected one is more pronounced there than in hetero-bilayers. More phonons are
indeed transferred to the top layer, if it is the same as the bottom one. Parts of these phonons are transmitted back to the
bottom layer, the rest is reflected back to its origin. Note that as for the PO configurations, the simulation results in Table~\ref{tab:others3} do not depend on the length of the overlap region.
This clearly indicates that stacking two TMD layers on top of each other, even if they weakly interact and a direct path exists between both electrodes, can have a profound impact on the thermal properties of such assemblies. 


\section{Conclusions and Outlook}\label{SEC:conclusion}
We explored the thermal transport properties of homo- and hetero-bilayer vdWMs made of two TMDs with total and partial overlap in the middle. To do that, we developed a suitable quantum transport approach where the simulation domain is created from structures with a full overlap of both TMDs and atoms are removed on both extremities, together with the corresponding entries of the dynamical matrix.
We demonstrated that in structures with full TMD overlap the inter-layer interactions marginally contribute to the thermal current, while their existence is crucial to enable the transfer of phonons from one layer to the other in case of partial overlap. For large overlap lengths, we showed that the ballistic thermal current flowing through partially overlapping channels is insensitive to the variation of this parameter. Partial overlap systems act as filters that reduce the ballistic thermal current by approximately 70\% for homo- and 95\%  for hetero-bilayers with respect to the thermal current of a channel with a full overlap of the constituting 2-D materials. Insights into the energy dependence of the simulated transmission functions highlighted the larger contribution of acoustic phonons to the thermal current. Moreover we proved that, as first approximation, the transmission function of the partially overlapping channel can be obtained by uniformly rescaling the transmission function of the corresponding structure with full overlap, when identical TMDs are considered. 
With different TMDs, the high energy phonon are filtered out, only the acoustic branches being transferred from one layer to the other.

It should be noted that a broadening of the phonon branches larger than the inter-band spacing induces anharmonic phonon interactions. The latter connect phonon modes that are independent from each other in the ballistic limit of transport. As the phonon dispersion of the vdWMs investigated in this work exhibit many closely spaced phonon bands, especially at low frequencies, anharmonic interactions could impact our results. However, we believe that this effect is not strong enough to compensate for the phonon dispersion mismatches observed in hetero-bilayer vdWMs.

While the proposed approach works very well for vdWMs with a weak inter-layer coupling, it might break down when the involved TMDs strongly interact with each other. Such case would require the introduction of a supercell, that exactly correspond to the simulation domain. To create such structure, DFT codes relying on localized basis set, e.g. CP2K \cite{cp2k}, are probably better suited than plane-waves, due to their linear-scale algorithms
\section*{Acknowledgements}
This research was supported by the NCCR MARVEL, funded by the Swiss National Science Foundation (SNSF), by the grant n$^{o}$ 175479 (ABIME) from SNSF and by the Swiss National Supercomputing Center (CSCS) under project S876
\appendix
\section{Dynamical Matrix Construction}\label{Sec:DevDM-general}
\begin{figure}
    \centering
    \includegraphics[width=8cm]{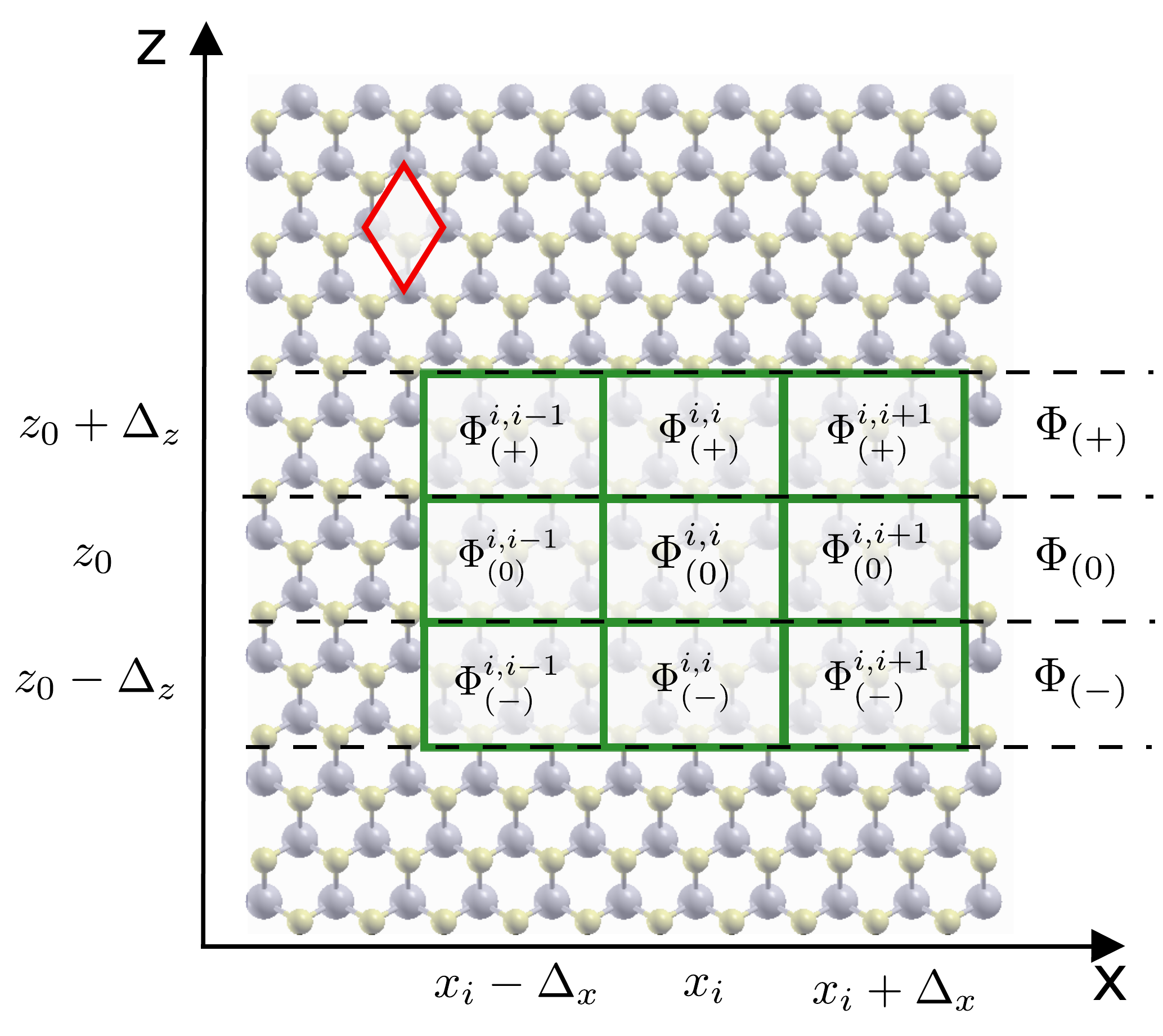}
    \caption{Top view of a MoS$_2$ structure, where \textit{x} denotes the transport direction, \textit{y} the direction of confinement and \textit{z}  is assumed to be periodic and modeled through a set of $q_z$ points. The dynamical matrices $\Phi(q_z)$ is computed by extracting all $\Phi^{i,j}_{(\alpha)}$ blocks required to construct the matrix of  Eq.~(\ref{device_DM}). The red area marks the primitive hexagonal cell, while the green box the orthorhombic unit cells.}
    \label{fig:PhiAtoms}
\end{figure}

In this section the process of assembling the dynamical matrix in Eq.~(\ref{Phi_eq}) is presented.
First, the unitary cell of the structure of interest must be determined and the atomic positions relaxed at high precision. By periodically replicating this cell, a larger supercell can be built, which is necessary to compute the interatomic force constants of the system. Since these interactions rapidly decay with the distance between atoms, a cutoff $r_{cut}$ is introduced, such that all interactions extending beyond it are neglected. It is important to note that choosing a (too) small $r_{cut}$ usually leads to inaccurate phonon dispersions. The determination of $r_{cut}$ is thus performed analysing the variation of the phonon band structure as a function of this parameter and by selecting the smallest $r_{cut}$ value that causes an average relative error below a predefined criterion, 10\% in our case. Once an appropriate $r_{cut}$ has been set an orthorhombic unit cell is constructed to allow for transport calculations.
 Its dimensions are chosen so that all interactions are confined within two neighboring cells at most. A graphical representation of this geometrical condition is depicted in Fig.~\ref{fig:PhiAtoms}, where the orthorhombic nearest neighbor cells are marked by green boxes. The exclusion of second (and more) nearest neighbor cells interactions is crucial to enable fast computation of the open boundary condition.
Along $z$ the periodicity is modeled by a $q_z$-dependence of the dynamical matrix 
 \begin{equation}\label{phialphaall}
  \Phi(q_z) = \Phi_{(0)} + \Phi_{(+)}e^{iq_z \Delta_z} + \Phi_{(-)}e^{-iq_z \Delta_z},   
 \end{equation}
where $\Delta_z$ is the width of the orthorhombic cell along $z$, $\Phi_{(0)}$ contains all the interactions within the central unit cell at $z=z_0$, while $\Phi_{(+)}$($\Phi_{(-)}$) describes the connections to the periodic replica situated at $z\,=\,z_0+\Delta_z$($z\,=\,z_0-\Delta_z$). Due to the choice of $r_{cut}$, all matrices in Eq.~(\ref{phialphaall}) have a block-tridiagonal structure and are made of $\Phi^{i,j}_{(\alpha)}$ blocks, where $\alpha= (0),(+),(-)$, $j=i-1,i,i+1$, and $i$ represents one of the $N_x$ blocks of width $\Delta_x$ along the transport direction $x$:
\begin{equation}
    \Phi_{(\alpha)}=      \begin{bmatrix}
    \Phi^{1,1}_{(\alpha)} & \Phi^{1,2}_{(\alpha)} &  &  &   \\
    \Phi^{2,1}_{(\alpha)} & \Phi^{2,2}_{(\alpha)} & \Phi^{2,3}_{(\alpha)} & &  \\
    \ddots& \ddots & \ddots&\\
     & \Phi^{i,i-1}_{(\alpha)} & \Phi^{i,i}_{(\alpha)} & \Phi^{i,i+1}_{(\alpha)} &  \\
     & \ddots& \ddots & \ddots&\\
     && \Phi^{N-1,N-2}_{(\alpha)} &\Phi^{N-1,N-1}_{(\alpha)}&\Phi^{N-1,N}_{(\alpha)}\\
      &   &  & \Phi^{N-1,N}_{(\alpha)} & \Phi^{N,N}_{(\alpha)} \\
    \end{bmatrix}.
    \label{device_DM}
\end{equation}
The size of each $\Phi^{i,j}_{(\alpha)}$ block is $3Na^i\times3Na^j$, where $3Na^i$ is the number of atoms in the orthorhombic unit cell located at $x=x_i$

{}


\section{Creation of the vdWM Dynamical Matrix}\label{Sec:inhomog-DDM}
\begin{figure}[h!]
    \centering
    \includegraphics[width=\linewidth]{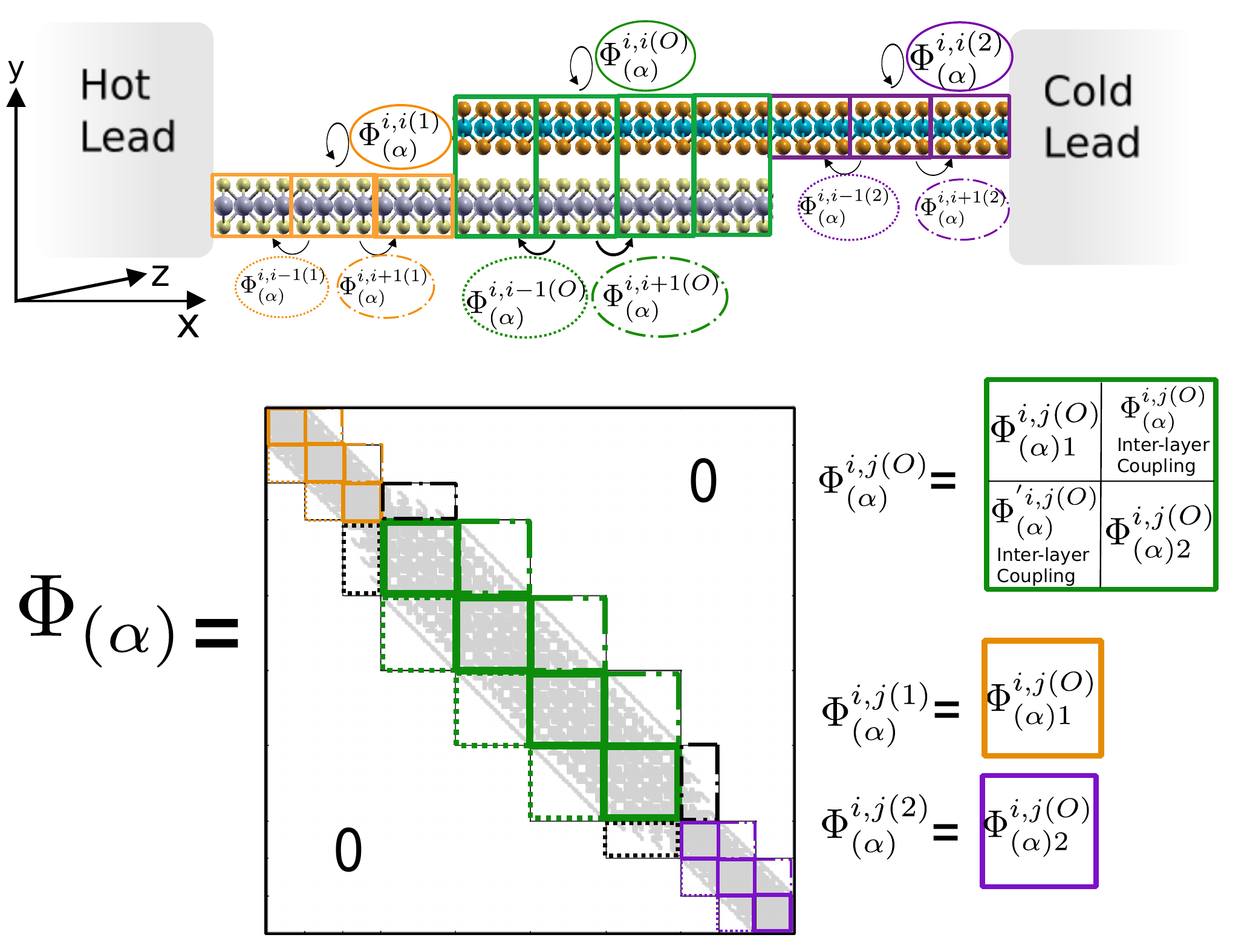}
    
    \caption{(\textit{Top}) Schematics of TMD monolayers stacked on top of each other with a partial overlap region in the middle. The on-site ($\Phi_{i,i}$) and nearest-neighbor ($\Phi_{i,i\pm1}$) blocks of the dynamical matrix are indicated. (\textit{Bottom}) Sparsity pattern of the corresponding device dynamical matrix. The blocks are colored according to their position in the device structure from the top panel.\label{fig:PO_matrix}}
\end{figure}

We are interested in modeling structures composed of TMD monolayers stacked on the top pf each other with a partial overlap region in the middle as shown  in  Fig.~\ref{fig:vdWHs}. In the case of a total overlap, the $\Phi^{i,j}_{(\alpha)}$ in Eq.~(\ref{device_DM}) are the same from one extremity of the vdWM to the other, regardless of the fact that both TMDs are identical or not. When instead, the two monolayers only partial overlap in the middle, three distinct regions can be identified. A Region1(2), where only TMD1(2) is present and an Overlap Region, where the two layers are stacked on top of each other, as illustrated in Fig.~\ref{fig:PO_matrix}.
There, the green cell in the Overlap Region corresponds to the one of the total overlap case is the one that was used to compute the interatomic force constants with DFT. The orange(purple) cell contains only the bottom(top) TMD, the top(bottom) layer and the corresponding entries of the dynamical matrix having been removed. The validity of this approach was demonstrated in Section~\ref{sec:effect-interlayer}, where it was shown that the inter-layer coupling does not strongly affect the properties of the individual layers. Concretely, in the overlap region, each  $\Phi^{i,j(O)}_{(\alpha)}$ block is made of four sub-blocks, the diagonal ones connecting two unit cells of the same TMD with each other, the off-diagonal blocks coupling the two TMDs together. To construct the dynamical matrix of Region1(2), the diagonal blocks of  $\Phi^{i,j(O)}_{(\alpha)}$ are extracted and labeled $\Phi^{i,j(1)}_{(\alpha)}$($\Phi^{i,j(2)}_{(\alpha)}$). The whole process is described in details in  Fig.~\ref{fig:PO_matrix}.
\newpage
\bibliography{references.bib} 
\end{document}